# Observation of a Dynamic Crossover in RNA Hydration Water which Triggers the Glass Transition in the Biopolymer

Xiang-qiang Chu[†], Emiliano Fratini[‡], Piero Baglioni[‡], Antonio Faraone[§], and Sow-Hsin Chen[†,*]

*Department of Nuclear Science and Engineering, Massachusetts Institute of Technology, Cambridge, MA 02139,*

*Department of Chemistry and CSGI, University of Florence, Florence, Italy*

*NIST Center for Neutron Research, Gaithersburg, MD 20899-8562, and Department of Material Science and Engineering, University of Maryland, College Park, MD 20742*

RECEIVED DATE (automatically inserted by publisher); E-mail: sowhsin@mit.edu

It is known that many proteins exhibit sharp slowing down of their functions (kinetics of biochemical reactions) at temperatures within the interval T~240-200 K. [1,2] It has also been shown experimentally[3-5] and computationally[6,7] that there is a sharp increase of the mean-squared atomic displacement $<x^2>$ in protein at about $T_C$ = 220 K, suggesting a dynamic transition (the so-called glass transition) in proteins at this temperature. There is strong evidence that this glass transition is solvent induced, since the hydration water of the protein also shows some kind of dynamic transition at a similar temperature.[8-10] In our previous research,[11] it was demonstrated that this dynamic transition of hydration water on lysozyme protein was, in fact, a fragile-to-strong dynamic crossover (FSC) at 220 K.

More recently, we also showed that hydration water on B-DNA exhibits a similar FSC at $T_L$ = 222 K.[12] To assess whether this glass transition temperature $T_C$ of proteins is universal for other kinds of bio-polymers, experiments[13,14] and computations[15] were performed on these other biopolymers. However, in these experimental works, there were no explicit demonstrations of the FSC phenomenon of the bio-molecules' hydration water.

This paper presents a new evidence of the universality of $T_C$. From the measurement of mean-square atomic displacement (MSD) of H-atoms in RNA and its hydration water, the similarity of the dynamic transition in RNA and its hydration water suggests that the dynamic transition in RNA is also induced by its hydration water, despite differences in the architecture and chemical backbones of proteins, DNA and RNA.

Using high-resolution quasi-elastic neutron scattering (QENS) spectroscopy, we demonstrate decisively that there is a sharp FSC temperature of the hydration water on RNA at $T_L$=220 K. The change of mobility of the hydration water molecules across $T_L$ drives the dynamic transition in RNA which happens at the same temperature. The coincidence of the crossover temperature in proteins,[11] DNA,[12] tRNA[14] and RNA reinforce the plausibility that the glass transitions are not the intrinsic properties of the bio-molecules themselves but are imposed by the hydration water on their surfaces.

Ribonucleic acid from torula yeast (RNA) was obtained from Sigma (R6875, batch number 021k7063) and used without further purification. The sample was extensively lyophilized to remove any water left. The dried RNA powder was then hydrated isopiestically at 5°C by exposing it to water vapor in equilibrium with a $NaClO_3$ saturated water solution placed in a closed chamber (relative humidity, RH =75%). The final hydration level was determined by thermo-gravimetric analysis and also confirmed by directly measuring the weight of absorbed water. A second sample was then prepared using $D_2O$ in order to subtract out the incoherent signal from hydrogen atoms of the RNA. Both hydrated samples had the same water or heavy water/dry RNA molar ratio. Differential scanning calorimetry analysis was performed in order to detect the absence of any feature that could be associated with the presence of bulk-like water.

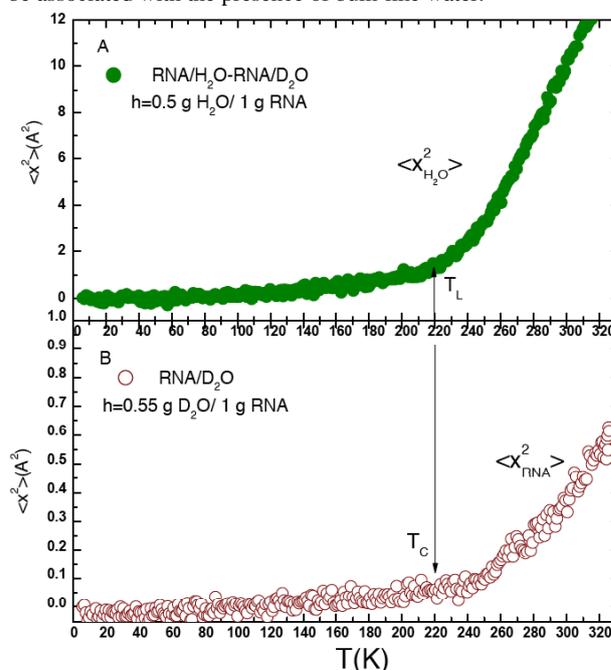

**Figure 1.** The mean-squared atomic displacement (MSD) averaged over all the hydrogen atoms, $<x^2>$, extracted from Debye-Waller factor measured by an elastic scan with resolution of 0.8 μeV, as a function of temperature for the $H_2O$ hydrated and $D_2O$ hydrated RNA samples. Panel A shows the data processed from the difference between the two, which gives MSD of the H-atoms in the hydration water. Panel B shows the data processed from the latter, which gives MSD of H-atoms in RNA. One can clearly see from the two panels, moderately sharp transition of slopes at around 220 K, indicating that the dynamic crossover temperatures of the RNA ($T_C$) and the hydration water ($T_L$) are approximately the same.

Neutron scattering measurements were done on the High-Flux Back-Scattering spectrometer at NIST-CNR with a resolution of

[†] Massachusetts Institute of Technology.
[‡] University of Florence.
[§] NIST-CNR and University of Maryland.

0.8 μeV and a dynamic range of ±11μeV. The elastic scans were measured at a heating/cooling rate of 0.75 K/min. It is used to deduce the mean-squared hydrogen atom displacement $<x^2>$, shown in Fig. 1A (from the difference signal of the two samples) and Fig. 1B (from the $D_2O$ hydrated sample).

Taking the difference of the QENS measurements of the two samples gives the Fourier transform of the Intermediate Scattering Function (ISF) of the hydrogen atoms, $F_H(Q,t)$, of water molecules on the surface of RNA. The Q-independent average translational relaxation time $\langle \tau_T \rangle$ is obtained from the difference QENS data by using the Relaxing Cage Model (RCM) for the hydration water, $F_{H_2O}(Q,t)$, as described in detail in reference[11].

Figure 1A shows the MSD of the hydration water molecule, $\langle x^2_{H_2O} \rangle$, in the observational time interval of about 2 ns (corresponding to the energy resolution of 0.8 μeV). Figure 1B shows the MSD of hydrogen atoms in the RNA molecule. From these two figures, one can conclude that dynamic transition temperature of the hydration water (denoted as $T_L$) and the glass transition temperature of the RNA molecule (denoted as $T_C$) are approximately the same (within the error bars of the kink positions). Figure 2 illustrates the fact that the normalized QENS peaks of the hydration water as a function of temperature qualitatively show that there is some kind of crossover temperature existing at around 220 K, judging from the variation of their peak height and peak width. However, a much more sharper definition of this dynamic crossover temperature $T_L$ can be obtained from RCM analysis of the difference spectra of the two hydrated samples. Figure 3 shows the $\log\langle \tau_T \rangle$ vs. 1/T plot of the average translation relaxation time.

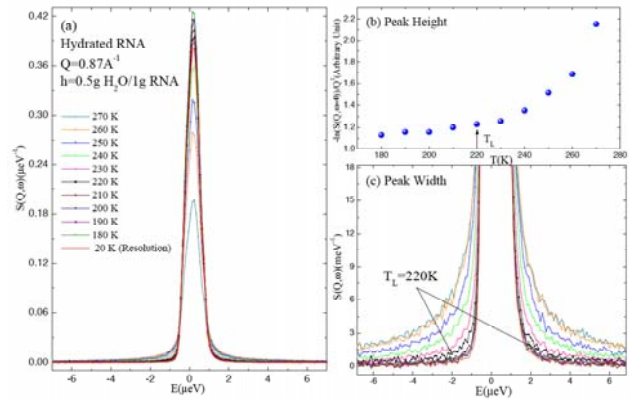

**Figure 2.** Measured difference QENS spectra at various temperatures T. Panel (a) shows the normalized QENS spectra at Q = 0.87Å$^{-1}$. Panel (b) shows the T dependence of the logarithm of the peak heights, which is linearly related to the MSD of the H-atoms of the hydration water. Panel (c) shows the wing of the peaks, from which we extract the average relaxation time $\langle \tau_T \rangle$. Both peak heights and widths indicate the existence of a well-defined crossover temperature $T_L$ for the hydration water.

At high temperatures, above 220 K, $\langle \tau_T \rangle$ obeys a Vogel-Fulcher-Tammann (VFT) law, namely, $\langle \tau_T \rangle = \tau_0 \exp[DT_0/(T-T_0)]$, where D is a dimensionless parameter providing the measure of fragility and $T_0$ is the ideal glass transition temperature. Below 220 K, the $\langle \tau_T \rangle$ switches to an Arrhenius behavior, which is $\langle \tau_T \rangle = \tau_0 \exp(E_A/RT)$, where $E_A$, is the activation energy for the relaxation process and R is the gas constant. This dynamic crossover from the super-Arrhenius to the Arrhenius behaviors is cusp-like and thus it sharply defines the crossover temperature to be $T_L$=220 K, much more accurately than that indicated by the MSD $\langle x^2_{H_2O} \rangle$, shown in Figure 1A.

We have shown in this paper a clear experimental evidence for the existence of a super-Arrhenius to Arrhenius dynamic crossover ($T_L$) in RNA hydration water at 220 K and proposed a plausible reason that the dynamic crossover of the hydration water triggers the onset of the glass transition at the same temperature ($T_C$) in RNA. We have shown before that above $T_L$ the structure of hydration water is predominately in its high-density form (HDL), which is more fluid. But below $T_L$, it transforms predominately to the low-density form (LDL), which is less fluid. Thus this abrupt change in the mobility of hydration water apparently induces the change in the energy landscape of RNA, which cause the dynamic transition or the glass transition in the biopolymer.

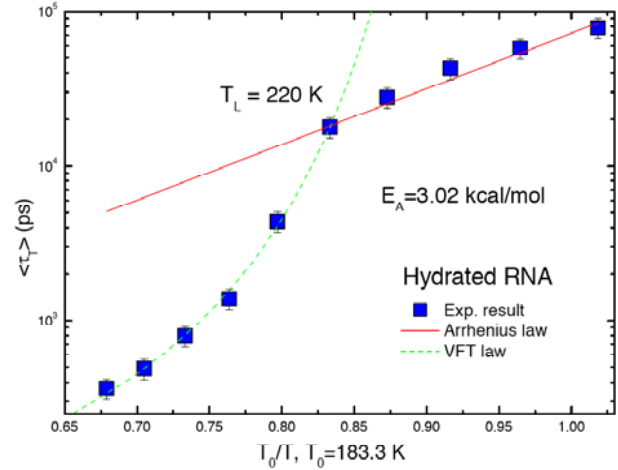

**Figure 3.** The extracted average translational relaxation time $\langle \tau_T \rangle$ from fitting of the quasi-elastic spectra by the relaxing cage model plotted in a log scale against 1/T. There is a clear evidence of a well-defined cusp-like dynamic crossover behavior occurring at $T_L$ = 220 K. The dashed line represents fitted curves using the VFT law, while the solid line is the fitting according to the Arrhenius law.

Acknowledgment. The research at MIT is supported by DOE Grants DEFG02-90ER45429 and 2113-MIT-DOE-591. E.F. and P.B. acknowledge CSGI (Florence, Italy) for partial financial support. This work utilized facilities supported in part by the National Science Foundation under Agreement No.DMR-0454672. Technical support in measurements from V. Garcia-Sakai at NCNR/NIST is greatly appreciated. We benefited from affiliation with EU-Marie-Curie Research and Training Network on Arrested Matter.